# Instruction Data Generation and Unsupervised Adaptation for Speech Language Models

*Vahid Noroozi[1], Zhehuai Chen[1], Somshubra Majumdar[1], Steve Huang[1],*
*Jagadeesh Balam[1], Boris Ginsburg[1]*

[1]NVIDIA, USA

{vnoroozi,zhehuaic,smajumdar,heh,jbalam,bginsburg}@nvidia.com

## Abstract

In this paper, we propose three methods for generating synthetic samples to train and evaluate multimodal large language models capable of processing both text and speech inputs. Addressing the scarcity of samples containing both modalities, synthetic data generation emerges as a crucial strategy to enhance the performance of such systems and facilitate the modeling of cross-modal relationships between the speech and text domains. Our process employs large language models to generate textual components and text-to-speech systems to generate speech components. The proposed methods offer a practical and effective means to expand the training dataset for these models. Experimental results show progress in achieving an integrated understanding of text and speech. We also highlight the potential of using unlabeled speech data to generate synthetic samples comparable in quality to those with available transcriptions, enabling the expansion of these models to more languages.

## 1. Introduction

Speech models are traditionally trained and fine-tuned using labeled data for single or multiple tasks. They achieve high performance in many targeted tasks but often exhibit limited generalization to out-of-domain tasks [1]. On the other hand, Large Language Models (LLMs) have demonstrated impressive text understanding and a wide range of emergent capabilities through training on predominantly unsupervised and unlabeled text data [2]. The success of LLMs has paved the way for the development of speech-language models capable of processing both text and speech inputs. These models are expected to leverage the text-understanding prowess of LLMs while comprehending speech inputs [3, 4, 5, 6, 7]. One major challenge in training such multimodal models is the scarcity of sufficient labeled data, necessitating paired speech and text samples. To address this challenge, we propose that the synthetic generation of data by utilizing the text-understanding capabilities of LLMs or text-to-speech systems (TTS) can significantly contribute to expanding the dataset required for modeling both text and speech inputs.

In this paper, we present and evaluate three strategies for generating instruction data to train speech language models:

**Synthetic Speech Instruction Data with TTS Systems**: Text-to-speech (TTS) systems are being used to generate the speech component of samples from a textual dataset. This approach is applied to a question-answering dataset, where the model must comprehend both textual and speech parts of the input, as well as their interrelation, to successfully answer the question. This method is versatile, applicable to any textual or instructional dataset for pre-training and instruction tuning. However, it faces limitations in speech diversity and the availability of efficient TTS models for certain languages, challenging the extension of speech models to a wider linguistic range.

**Text Generation from Labeled Speech Data**: We leverage the text-understanding capabilities of LLMs to generate textual content for labeled speech data. In this approach, we use LLMs to generate questions and answers based on the context provided by the transcription of speech data. By treating the speech segment as the context, we create triplets of question/context/answer, where the context is real speech, and the remaining elements are in textual form. This method significantly expands data generation with real and diverse speech data. However, the quality of the text generated by LLMs may not always be optimal, so we explore using LLMs as judges to evaluate and filter the quality of the generated samples.

**Text Generation from Pseudo-Labels**: We investigate the use of pseudo-labels generated by automatic speech recognition (ASR) systems to generate textual parts for speech data. This approach addresses the limitation of relying solely on labeled speech samples and enhances the model's generalization to more languages, especially those with limited resources. Our experiments demonstrate that pseudo-labels can be as effective as high-quality transcriptions, highlighting the potential of leveraging a vast amount of unlabeled speech data.

We have evaluated and studied the proposed approaches on a speech language model and showed that synthetic data generation is a promising avenue for achieving joint speech and text understanding, unlocking emergent capabilities, and capitalizing on the text-understanding prowess of LLMs. We also show with examples that the trained model's capabilities are not limited to the tasks trained for and it can show some degree of generalization to other tasks. All the experiments and model training are done with NeMo [8] toolkit[1].

## 2. Previous Works

TTS systems have been used to generate synthetic speech instruction data in [9] primarily for evaluation. They have used TTS models to generate speech for the question-answering dataset Trivia-QA [10] and evaluated their proposed model on the synthetic QA benchmark. Speech-language model [11] was proposed to bridge the gap between speech and text foundation models. They used TTS to convert the context part of the instruction samples from the Alpaca dataset [12] and used it as an instruction-tuning dataset to fine-tune their model. They also tested their model's capability on a Spoken QA dataset generated with TTS without performing or reporting any subjective evaluations.

Other models such as SALMONN [6], COSMIC [13], and LibriSQA [14] have employed ChatGPT-3.5 [2] to generate questions and answers for speech datasets as part of their training data. However, these studies provide limited details on their methodologies, and some [6] have not conclusively demonstrated the effectiveness of this approach for their specific goals and tasks. Additionally, previous work has not extensively discussed the impact of filtering out low-quality synthetic samples from the dataset or the use of unlabeled speech data in their models.

---

[1]https://github.com/NVIDIA/NeMo

## 3. Instruction Data Generation

In this section, we present three distinct methodologies for generating instruction data that encompass both text and speech, facilitating the training and evaluation of models capable of comprehending both modalities as input.

### 3.1. Synthesized from textual data

In this approach, we transform textual samples into paired samples comprising both text and speech. We utilize a text-to-speech (TTS) model to convert segments of text into speech, resulting in paired samples. To enhance the model's comprehension of the interdependent relationship between the textual and speech components, we employ samples that require an understanding of both elements to formulate a response. In our experimental setup, we utilize public question-answering datasets, converting the context portion of the samples into speech. To increase the diversity of the speech, we employed a multi-speaker TTS system to generate speech for the context component. We randomly selected a speaker for each sample and generating the speech parts offline prior to training the model.

### 3.2. Synthesized from labeled speech

A limitation of generating speech parts using TTS is the restricted diversity of the generated speech and the limited availability of TTS systems for various languages. To expand the synthetic generation of samples for speech-language models, we propose an approach that leverages real speech samples. We use LLMs to generate textual components for speech samples that have available transcriptions. We use a prompt with a few-shot examples to prime the LLM to generate a question-and-answer pair for a textual context [15, 16]. The context for each sample is the transcription of the speech sample. The generated question and answer, along with the speech sample, form a new text-speech paired sample (see Fig. 1).

Filtering synthetic samples generated by LLMs is vital for the efficiency and accuracy of the training process. To filter out low-quality generated samples we used LLM with filtering prompt (see Fig. 2).

### 3.3. Synthesized from unlabeled speech

The generation of synthetic QA samples from labeled speech data is inherently limited by the availability of labeled speech samples, which may be scarce for many languages. Even in languages with abundant resources, utilizing unlabeled speech samples can offer significant advantages. In this approach, we transcribe speech samples using an automatic speech recognition (ASR) model for the respective language and use the resulting pseudo-labels to generate synthetic paired samples. Our experiments demonstrate that pseudo-labels can be as effective as high-quality transcriptions, and that the transcription does not need to be perfectly accurate to produce useful samples.

[2]https://chat.openai.com

## 4. Experiments

### 4.1. Experimental Settings

In our evaluations, we primarily test our method using the SALM model [7], an open-source speech-language model proficient in speech recognition, word boosting, and translations. This model is designed to leverage a pre-trained and instruction-fine-tuned LLM by conditioning it on paired speech and text prompts to generate textual outputs for various speech tasks. It is trained using supervised speech instruction tuning data [7] with LoRA [17]. The encoder is initialized from the 110M parameter FastConformer-large ASR model [3]. By default, we use TinyLlama-1.1B-chat [18] as the frozen LLM backbone.

Our baseline model is trained using the LibriSpeech dataset [19]. In all experiments, we augment the original training dataset with the proposed synthetic samples (upsampled 3X) to enhance the model's capabilities. To demonstrate that our method is model-agnostic, we explore alternative architectures and scale up the LLM backbones to Llama-2-7b-chat [20] in subsequent sections.

Following the training protocol of SALM [7], we train the model with a global batch size of 512 for three epochs, using the Adam optimizer with a learning rate of 1e-4 and a weight decay of 0.001, on 32 V100 GPUs. By default, we employ greedy decoding for inference. For 1B LLMs, we use a 256-dimensional LoRA, and for 7B LLMs, we use a 32-dimensional LoRA. We train 7B LLMs with a tensor-model-parallel size of 4.

### 4.2. Datasets

We evaluate the performance of our methods on two proposed synthetic datasets: Speech-MSMARCO and SpokenQA-LS. Speech-MSMARCO is derived from the MS-MARCO (MicroSoft MAchine Reading COmprehension) dataset [21], a textual question-answering dataset with samples in text format. In the original dataset, each sample consists of a question with multiple contexts, with the answer potentially present in one of the contexts. We modified the original dataset to create a set of triplets consisting of a question, context, and answer. The context containing the answer was paired with the corresponding question and answer to form a triplet. We employed a multi-speaker TTS system, composed from FastPitch [22] with HiFiGAN vocoder [23] to generate speech for the context part while retaining the question and answer in text format[4]. We randomly subsampled the generated dataset, retaining 109K samples with synthesized durations under 20 seconds for the training set and 1,000 samples each for development and testing. The training subset of the new synthetic dataset, named Speech-MSMARCO, was then combined with the original training data to create a joint training dataset.

SpokenQA-LS is a larger dataset with 2.8 million samples in a similar format to Speech-MSMARCO. However, it features real speech data sourced from the LibriSpeech (LS) dataset, with the questions and answers generated by LLMs. We utilized LLMs to generate question-and-answer pairs for the triplets, with the transcription text of speech samples serving as their contexts. In some instances, the outputs generated by the LLMs did not adhere to the prescribed output format, leading to their rejection as failed generations. We generated from LS dataset approximately 2.8 million triplets using the Mixtral 8X7B model [24]. During decoding, we set the temperature to 1.0 and the top-p value to

[3]https://catalog.ngc.nvidia.com/orgs/nvidia/teams/nemo/models/stt_en_fastconformer_transducer_large
[4]https://catalog.ngc.nvidia.com/orgs/nvidia/teams/nemo/models/tts_en_multispeaker_fastpitchhifigan

> **Generation Prompt**
>
> I want you to act as an expert Transcriber and Analyst. You are asked to come up with a set of 20 diverse questions and their corresponding answers that are related to the given text. These task instructions will be given to a GPT model and we will evaluate the GPT model for completing the instructions.
>
> Here are the requirements:
> 1. Try not to repeat the verb for each instruction to maximize diversity
> · · ·
> 6. The Output should be an appropriate response to the instruction and the #Given Transcript#.
>
> #Examples#
> {examples}
> #Given Transcript#
> {transcript}
> List of 20 questions - answer pairs that are related to the #Given Transcript#:

> **Few-Shot Example Format**
>
> 1. Instruction: Can you summarize the main idea of the transcript in a single sentence?
> 1. Corresponding Transcript: i never expected this but i'll do for you mister finn just as i ever did in the old days and it was i that was sorry when i heard of the poor young lady's death so i was mister finn well then i won't mention her name never again
> 1. Output: An old friend of Mister Finn expresses their condolences for the death of a young woman and promises not to mention her again.
> 2. Instruction: · · ·

Figure 1: *Prompt format for Q&A instruction generation, with few-Shot in-context learning.*

Table 1: *Comparative performance of the various training datasets. Performance of the models on the ASR task is evaluated on the test-other set of the LibriSpeech dataset. The performance of the QA task is measured with ROUGE-L on test sets of the Speech-MSMARCO (SP-MSMARCO) and SpokenQA-LS (SQA-LS).*

| Datasets Trained on | test-other WER% | SP-MSMARCO ROUGE-L | SQA-LS ROUGE-L |
|---|---|---|---|
| LibriSpeech | 5.6 | 0.36 | 0.24 |
| + SP-MSMARCO | 5.5 | 0.50 | 0.24 |
| + SQA-LS | 5.7 | 0.36 | 0.41 |
| + SQA-LS-filtered | 5.6 | 0.36 | **0.42** |
| + SQA-LS + SP-MSMARCO | **5.4** | **0.57** | 0.42 |
| + Cascaded MSMARCO LLM | NA | **0.57** | NA |

0.95. The newly generated dataset is named SpokenQA-LS, with 2,000 samples randomly selected as the test set and the remaining samples used for training. For the filtering step, the same LLM was employed, resulting in approximately 1.8 million final samples, referred to as SpokenQA-LS-Filtered in the results.

### 4.3. Synthetic speech instruction data with TTS Systems

Models were evaluated on the test set of Speech-MSMARCO using the ROUGE-L metric [25], and the results are reported in Table 1. Additionally, the Word Error Rate (WER%) of the models on the test-other set of LibriSpeech (LS) was reported to demonstrate the models' ability to maintain their original ASR capabilities. The results indicate that the model trained with Speech-MSMARCO significantly improved performance on the test set while preserving its ASR capabilities, demonstrating the

> **Generation Prompt**
>
> You are an accurate AI assistant which is designed to evaluate the quality and accuracy of context/question/answer triplets. Each question should be relevant to the given context, and it should be a question which can get answered based on the context. Also the answer should also be correct and grounded based on the given context. Please evaluate the quality of both the question and answer and reject or accept a triplet.
>
> You may follow the following guidelines:
> 1. The question should be related to the context and can get answered from the context.
> · · ·
> Please first provide a brief reasoning you used to derive the decision, and then write "ACCEPT" or "REJECT" in the last line. Here are 3 examples of how to evaluate

> **Few-Shot Example Format**
>
> Example 1: #Context#
> mister thornhill having been there that day to inform them that their journey to town was entirely over the two ladies having heard reports of us from some malicious person about us were that day set out for london
> #Question#
> In the given transcript, who is "mister thornhill"?
> #Answer#
> Mister Thornhill is the person who went to inform the two ladies that their journey to town was completely over.
> #Evaluation#
> The question is relevant to the context. The answer is a correct reply to the asked question, and it is grounded based on the context.
> ACCEPT
> Example2: · · ·

Figure 2: *Prompt format for filtering Q&A triplets with few-shot in-context learning.*

effectiveness of the proposed approach in enabling the model to learn the inter-modality relationship between the text and speech components of the inputs.

We also evaluated the model in cascaded mode where TinyLlama-1.1B-chat is fine-tuned with LORA on the textual MSMARCO and the QA context is transcribed with the first row model and then given to the LLM model as input. The cascaded model shows similar performance to the models fine-tuned jointly as expected.

### 4.4. How to use real speech?

In this section, we assess the efficacy of employing LLMs to generate textual components of triplets for speech samples using our proposed method in this section. The performance of the base model trained with variants of SpokenQA-LS is shown in Table 1. We also trained a model using a combination of the Speech-MSMARCO and SpokenQA-LS training sets. Training on SpokenQA-LS improved accuracy on its own test set but did not significantly enhance performance on Speech-MSMARCO. Conversely, the model trained on Speech-MSMARCO exhibited the opposite trend. The model trained on a combination of both datasets achieved the best results on both evaluation sets. These experiments demonstrate that the two approaches are complementary: while Speech-MSMARCO boasts high quality and diversity in its textual components, SpokenQA-LS offers greater diversity in the speech domain. Combining these approaches can enhance the model's robustness and accuracy across a broader range of scenarios.

**1st Turn**

**Audio Transcript**: I'm going to talk today about energy and climate.
**Question**: what's the main topic today?
**Answer**: Energy and climate are the main topics today.

**2nd Turn**

**Audio Transcript**: And that might seem a bit surprising, because my full-time work at the foundation is mostly about vaccines and seeds, about the things that we need to invent and deliver to help the poorest two billion live better lives.
**Question**: what's the main job of the speaker?
**Answer**: The speaker is talking about the main topics of today.
...

**6th Turn**

**Audio Transcript**: Also, the price of energy is very important to them.
**Question**: who is affected by the price of energy for most?
**Answer**: the poorest two billion people on the planet

Figure 3: *An example of a multi-turn conversation with multiple speech audios.*

We have also tried the model trained on the synthetic datasets in multi-turn conversations with multiple speech audios and it could handle the conversation. A partial example is given in Fig. 3 for a six-turn conversation. It is an evidence that the model has some degree of adaptability to other tasks and domains.

### 4.5. How to use unlabeled speech data?

We study the impact of high-quality transcriptions versus machine labels for speech data on the QA synthetic process. For this investigation, we utilize the test split of the SPGI dataset [26], which contains 38K speech utterances from the finance domain. We reserve 5K utterances as the test set and treat the remaining 33K as unlabeled data for the experiment. The 33K speech samples are transcribed using the baseline SALM model to generate *pseudo-labels*, and then the same LLMs from section 3.2 are employed to generate the questions and answers.

We also conduct a similar data generation process using *real-labels* as a reference for oracle performance. The WER evaluation is conducted on the reserved 5K utterances. ROUGE-L scores are reported on 4.5K utterances of synthetic and filtered question-and-answer data generated from the *real-labels* of the above-mentioned dataset. The results are presented in Table 2. The model trained with high-quality ground-truth labels exhibits similar QA performance to the one trained with pseudo-labels. There is still a gap on *SPGI-test WER* performance, which on one hand shows how noisy the pseudo-labels are; on the other hand can be improved by better semi-supervised learning methods in previous research [27]. Also we did not observe any degradation on ASR accuracy on the test-other set of LS when trained and adopted for SPGI. This finding suggests that the quality of the transcriptions does not need to be perfect, presenting an opportunity to utilize unlabeled speech data for generating synthetic data.

### 4.6. Is the proposed method model-agnostic?

To demonstrate that the the proposed method is model-agnostic, we explored alternative LLM backbones (see Table 3). In addition to the SALM architecture, we investigated a cross-attention variant of SALM [7], denoted as SALM-XATT. Unlike the standard SALM approach, which prepends speech prompts to text prompts as LLM inputs, SALM-XATT applies cross-attention between speech prompts and the original textual LLM inputs at every step before feeding them into the LLM. In this cross-attention mechanism, the query is the original LLM inputs, and the key and value are the speech prompts, similar to the LAS model design for end-to-end ASR [28]. A more detailed comparison between the two architectures can be found in [29]. We mixed the two data generation approaches as found to be the best in previous sections. As shown in the table, scaling up the LLM backbone consistently delivered additional benefits, while the proposed method remained effective across different Speech-LLM architectures and LLM backbones.

Table 2: *Performance comparison of datasets built with pseudo-labels vs real labels from SPGI dataset. Performance of the ASR task is evaluated on the SPGI-test dataset while the performance of the QA task is measured with ROUGE-L on test sets of the SpokenQA-SPGI (SQA-SPGI).*

| Datasets Trained on | SPGI Labels | SPGI-test WER% | SQA-SPGI ROUGE-L |
|---|---|---|---|
| LibriSpeech | - | 21.0 | 0.21 |
| + SPGI | pseudo | 18.0 | 0.21 |
| + SQA-SPGI | pseudo | 20.5 | 0.35 |
| + SPGI + SQA-SPGI | pseudo | 18.2 | **0.35** |
| + SPGI | real | **6.0** | 0.21 |
| + SQA-SPGI | real | 21.0 | **0.35** |
| + SPGI + S-SPGI | real | **6.0** | **0.35** |

Table 3: *Comparison of different architectures trained on the mixture of LS, SpokenQA-LS, and Speech-MSMARCO.*

| Model Arch. | Speech-MSMARCO ROUGE-L | SpokenQA ROUGE-L |
|---|---|---|
| SALM - 1B | 0.57 | 0.42 |
| SALM - 7B | 0.58 | 0.54 |
| SALM-XATT - 1B | 0.45 | 0.45 |
| SALM-XATT - 7B | 0.53 | 0.49 |

## 5. Conclusion and Limitations

In this paper, we introduced and explored various methods for generating synthetic samples that encompass both speech and text. These samples are instrumental in training multimodal speech-language models capable of processing both text and speech inputs. By leveraging these samples, models can jointly represent both modalities and grasp the relationship between them. Our methodologies utilize TTS systems and LLMs to synthesize either the speech or textual components of the samples. We demonstrated that unlabeled speech data could serve as a viable source for synthetic data generation, eliminating the necessity for high-quality transcriptions. Our experiments with a speech-language model validated the effectiveness of our proposed approaches in enhancing the model's multimodal understanding capabilities.

A limitation of our work is that our synthetic data primarily focuses on content, which prevents the model from learning all the acoustic features of speech. It would limit the capability of the model on extending its understanding to speech features like emotions. A future direction for this research is to extend it to more diverse domains, such as audio understanding or emotion recognition.